\documentclass[article]{IEEEtran}
\IEEEoverridecommandlockouts
\usepackage{amsmath,amssymb,amsfonts}
\usepackage{algorithmic}
\usepackage{graphicx}
\usepackage{textcomp}
\usepackage{xcolor}
\usepackage{authblk}
\def\BibTeX{{\rm B\kern-.05em{\sc i\kern-.025em b}\kern-.08em
    T\kern-.1667em\lower.7ex\hbox{E}\kern-.125emX}}

\usepackage{booktabs}
\usepackage{hyperref}
\usepackage{multirow}
\usepackage{tabularx}
\usepackage{array}

\usepackage{csquotes}  

\usepackage{orcidlink}
\begin{document}

\newcommand{\paperabbr}{GEC-RAG}

\title{
\paperabbr: Improving Generative Error Correction via Retrieval-Augmented Generation for Automatic Speech Recognition Systems
}



\author[1]{Amin Robatian}
\author[1]{Mohammad Hajipour}
\author[1]{Mohammad Reza Peyghan}
\author[2]{Fatemeh Rajabi}
\author[1]{Sajjad Amini
	\textsuperscript{*}\thanks{* Corresponding author: Sajjad Amini, s\_amini@sharif.edu}}
\author[1]{Shahrokh Ghaemmaghami}
\author[1]{Iman Gholampour}

\affil[1]{Electrical Engineering Department, Sharif University of Technology, Tehran, Iran}
\affil[2]{Department of Mathematics and Computer Science, Amirkabir University of Technology, Tehran, Iran}

\maketitle

\begin{abstract}
Automatic Speech Recognition (ASR) systems have demonstrated remarkable performance across various applications. However, limited data and the unique language features of specific domains, such as low-resource languages, significantly degrade their performance and lead to higher Word Error Rates (WER). In this study, we propose Generative Error Correction via Retrieval-Augmented Generation (GEC-RAG), a novel approach designed to improve ASR accuracy for low-resource domains, like Persian. Our approach treats the ASR system as a black-box, a common practice in cloud-based services, and proposes a Retrieval-Augmented Generation (RAG) approach within the In-Context Learning (ICL) scheme to enhance the quality of ASR predictions.
By constructing a knowledge base that pairs ASR predictions (1-best and 5-best hypotheses) with their corresponding ground truths, GEC-RAG retrieves lexically similar examples to the ASR transcription using the Term Frequency-Inverse Document Frequency (TF-IDF) measure. This process provides relevant error patterns of the system alongside the ASR transcription to the Generative Large Language Model (LLM), enabling targeted corrections.
Our results demonstrate that this strategy significantly reduces WER in Persian and highlights a potential for domain adaptation and low-resource scenarios.
This research underscores the effectiveness of using RAG in enhancing ASR systems without requiring direct model modification or fine-tuning, making it adaptable to any domain by simply updating the transcription knowledge base with domain-specific data.
\end{abstract}


\section{Introduction}
Automatic Speech Recognition (ASR) systems are designed to transform speech signals into sequences of words, supporting applications such as text-based communication and device control \cite{errattahi2018automatic}.
In recent advancements, ASR technology has seen significant progress, transitioning from traditional Hidden Markov Model (HMM)-based frameworks
\cite{huang2001spoken}
to cutting-edge end-to-end (E2E) models like wav2vec 2.0
\cite{baevski2020wav2vec}
and Whisper \cite{,radford2023robust}.
However, due to the lack of data in low-resource languages or specific domains, these systems face challenges and errors.
The ongoing presence of ASR errors has increased the need to develop techniques for automatically detecting and correcting errors in ASR systems \cite{errattahi2018automatic}.

Given the limitations of paired data for training ASR systems, the use of Language Models (LMs) trained on datasets several orders of magnitude larger has been investigated to improve ASR performance
\cite{toshniwal2018comparison,shin2019effective,guo2019spelling,shan2019component, hrinchuk2020correction,mcdermott2019density,wang2020asr,liu2021asr,dutta2022error,prabhavalkar2023end, ma2023n}.
These approaches are typically classified into two categories: language model fusion methods and Error Correction (EC) techniques as a post-processing step.
In this paper, we focus on methods that treat the ASR system as a black-box, an approach that is particularly significant given the increasing use of cloud-based ASR services through APIs. Therefore, we will concentrate on EC methods, which offer a practical solution for improving ASR accuracy without requiring modifications to the core ASR system.

The study by Shin et al. \cite{shin2019effective} employs the BERT model to rescore N-best list hypotheses generated by ASR systems, significantly improving recognition accuracy through enhanced contextual understanding. Similarly, Guo et al. \cite{guo2019spelling} propose a spelling correction model for end-to-end ASR systems, effectively addressing transcription and spelling errors as a post-processing step. Hrinchuk et al. \cite{hrinchuk2020correction} present a transformer-based sequence-to-sequence model for ASR error correction, demonstrating its ability to learn error patterns and generate refined transcriptions. Furthermore, Dutta et al. \cite{dutta2022error} showcase the effectiveness of BART in ASR error correction by utilizing diverse data augmentation techniques during fine-tuning, leading to significant WER reductions, with further improvements achieved through alignment and rescoring strategies. In addition, Ma et al. \cite{ma2023n} introduce an N-best T5 model fine-tuned from a pre-trained T5 language model, which leverages ASR N-best lists to improve transcription accuracy. By transferring knowledge from the language model and extracting richer information from the ASR decoding space, this method outperforms a strong Conformer-Transducer baseline.

Recent advancements in generative Large Language Models (LLMs) have introduced a transformative paradigm for Generative Error Correction (GER) in automatic speech recognition, aiming to predict accurate transcriptions from the decoded N-best hypotheses through the application of generative LLMs \cite{ma2023can,ma2024asr,li2024investigating,hu2024listen,li2024rag,pusateri2024retrieval,ghosh2024failing}. In the studies by Ma et al. \cite{ma2023can} and \cite{ma2024asr}, large language models, particularly ChatGPT, are leveraged for ASR error correction, showing that generative models can significantly enhance transcription accuracy by utilizing their deep language comprehension, without requiring task-specific fine-tuning. Li et al. \cite{li2024investigating} fine-tune a multilingual LLM to correct 1-best ASR hypotheses in over 100 languages, demonstrating that the approach not only boosts ASR performance but also improves error correction in low-resource languages via knowledge transfer between languages with shared writing systems. Hu et al. \cite{hu2024listen} introduce ClozeGER, an innovative ASR error correction paradigm using SpeechGPT, a multimodal LLM that incorporates source speech to refine correction accuracy, and reformulates GER as a cloze test to minimize redundancy and simplify the process. Li et al. \cite{li2024rag} present LA-RAG, a novel Retrieval-Augmented Generation (RAG) approach, which enhances ASR accuracy by integrating token-level speech data stores and a speech-to-speech retrieval mechanism. Pusateri et al.
\cite{pusateri2024retrieval} propose a RAG-like approach for correcting entity name errors in ASR systems, particularly for low-frequency entities. Their method involves indexing relevant entities in a vector database and utilizing an LLM to integrate the retrieved entities into the correction process, thereby enhancing ASR outputs.
\cite{ghosh2024failing} introduces DARAG, which fine-tunes LLaMA-2 on augmented datasets containing synthetic errors and incorporates named entity retrieval to improve ASR error correction.

In this study, we introduce Generative Error Correction via Retrieval-Augmented Generation (\paperabbr), a novel approach designed to enhance the performance of ASR systems, particularly for low-resource languages.
The key innovation of \paperabbr\
lies in leveraging a knowledge base that stores ground truth data alongside ASR predictions, including 1-best and 5-best hypotheses.
This knowledge base functions as a resource for retrieving examples based on lexical similarity, utilizing the
TF-IDF \cite{tfidf}
approach to identify relevant matches,
which are then used for In-Context Learning (ICL) with the GPT-4o model.
By explicitly exposing the model to the typical error patterns of the ASR system,
\paperabbr\
enables precise and effective error correction without requiring any modifications to the ASR system itself. This approach is particularly well-suited to the challenges of low-resource language scenarios, offering a scalable, adaptable, and efficient solution for reducing Word Error Rate (WER) while maintaining compatibility with black-box ASR services.
Our contributions are as follows:
\begin{itemize}
    \item 
    We propose a novel retrieval-based approach for ASR error correction that operates externally to the ASR system.
    
    \item
    We demonstrate the effectiveness of in-context learning with GPT-4o for correcting ASR errors in Persian, a low-resource language.
    
    \item
     We show how our method achieves significant improvements in WER and is particularly advantageous for domain adaptation.
    
\end{itemize}

The remainder of this paper is organized as follows:
Section \ref{sec:methodology} provides a comprehensive overview of the proposed methodology, detailing the design and construction of the knowledge base, the retrieval process in the RAG framework, and the integration of in-context learning using GPT-4o to address ASR errors effectively.
Section \ref{sec:Experiments} outlines the experimental setup, including the description of datasets, the evaluation metrics employed, and the experimental results. This section highlights the performance improvements achieved by
\paperabbr
, particularly in reducing WER for Persian, a low-resource language.
Section \ref{sec:conclusion} wraps up the paper by summarizing the key findings, highlighting the practical benefits of \paperabbr, and outlining directions for future research, including its potential extension to other low-resource languages and specialized domains.

\section{Methodology}
\label{sec:methodology}

In this section, we describe the methodology of
\paperabbr,
a novel approach designed to improve the accuracy of ASR systems by leveraging retrieval-augmented generation. The entire
\paperabbr\
process can be mathematically represented as:  

\begin{equation}
t = G(A(v), R(A(v), K))
\label{eq:RAG_EC}
\end{equation}

Where:
\begin{itemize}
    \item \(v\): Represents the voice input, which is the raw audio containing spoken language.
    \item \(A\): The ASR system \(A\) processes the voice input \(v\) and generates an initial transcription. This transcription may contain errors due to various factors like noise, accents, or misrecognition.
    \item \(K\): The knowledge base, comprising a collection of preprocessed examples and their transcripts. This serves as a repository of reference materials for retrieval.
    \item \(R\): The retriever function, which identifies and retrieves passages from \(K\) that are relevant to the query \(A(v)\) lexically.
    It ensures that the retrieved samples align closely with the surface structure and intent of 
    \(A(v)\).
    \item \(G\): The generator function, which takes the query \(A(v)\) and the retrieved passages \(R(A(v), D)\) as inputs. It produces the corrected response \(t\).
\end{itemize}

This mathematical representation encapsulates the core operations of 
\paperabbr: 
retrieval and generation.
By grounding the response \(t\) in surface-level data, the approach effectively enhances the accuracy and reliability of the final output.

\begin{figure}[htbp]
\centerline{\includegraphics[width=0.95\linewidth]{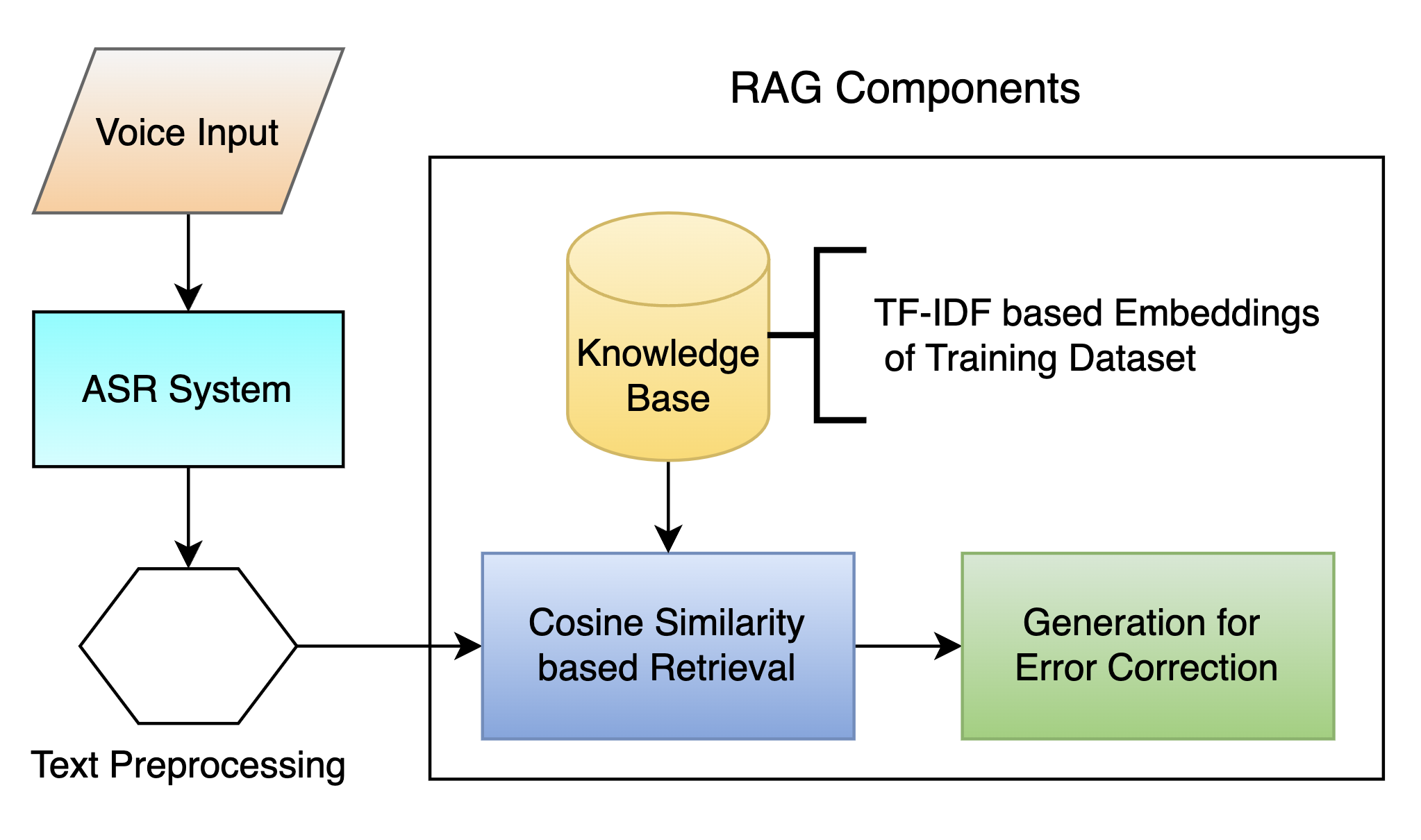}}
\caption{Workflow of the \paperabbr\ approach for error correction in ASR systems}
\label{fig1}
\end{figure}

\textbf{Figure \ref{fig1}} illustrates the workflow of \paperabbr, which is composed of four main stages based on the methodology explained: 1) Speech-to-Text conversion, 2) Knowledge base creation, 3) Retrieval of lexically similar samples, and 4) Generating corrected answers. Below, we detail each step of the pipeline.

\begin{table*}[htbp]
\centering
\caption{K-shot N-best Strategy for ASR Error Correction. Here, K and N are 5 and (1 or 5), respectively.}
\label{tab:my-table}
\resizebox{\textwidth}{!}{%
\begin{tabular}{|l|ll|}
\hline
\multicolumn{1}{|l|}{\textbf{Role}} & \multicolumn{2}{c|}{\textbf{Content}}                                                                                                                                                                                                                                                                            \\ \hline
\multirow{8}{*}{\textbf{System:}}   & \textbf{1-Best}                                                                                                                                                                                                                                                                                                                                                                                                                                    & \begin{tabular}[c]{@{}l@{}}You are a transcription error correction assistant and linguistics expert, specializing in improving transcriptions produced\\ by Automatic Speech Recognition (ASR) systems.\\ Your task is to identify and correct errors in transcriptions, with a focus on likely misinterpretations, such as homophones,\\ phonetically similar words, and contextually inappropriate word choices.\\ Use your linguistic expertise to analyze and enhance transcription accuracy while considering the intended meaning.\\ Provide only the correct transcription if needed, or repeat the text exactly if no correction is needed.\\ Do not include any explanations or additional commentary.\end{tabular}                                                                                                              \\
                                    &                                                                                                                                                                                                                                                 &                                                                                  \\
                                    & \textbf{N-Best}                                                                                                                                                     & \begin{tabular}[c]{@{}l@{}}You are a transcription error correction assistant and linguistics expert, specializing in improving transcriptions produced \\ by Automatic Speech Recognition (ASR) systems.\\ Your task is to perform error correction based on the top N outputs generated by the ASR system, which are listed in order \\ of their ASR posterior score.\\ Identify and correct errors in the transcriptions, focusing on likely misinterpretations such as homophones, phonetically similar words, \\ and contextually inappropriate word choices.\\ Analyze the linguistic context and provide the corrected ASR hypothesis directly without any explanations or additional commentary.\end{tabular}                                                                                                       \\ \hline
\textbf{User:}                      & \multicolumn{2}{l|}{\begin{tabular}[c]{@{}l@{}}Example1:\\ Predicted Transcriptions:\\ \textless{}hypothesis1\textgreater{}1st-Best Prediction \textless{}/hypothesis1\textgreater\\ \textless{}hypothesis2\textgreater{}2nd-Best Prediction \textless{}/hypothesis2\textgreater\\ ...\\ \textless{}hypothesisN\textgreater{}Nth-Best Prediction \textless{}/hypothesisN\textgreater\\ =\textgreater{}Correct Transcription: Label 1\\ Example2:\\ Predicted Transcriptions:\\ \textless{}hypothesis1\textgreater{}1st-Best Prediction \textless{}/hypothesis1\textgreater\\ \textless{}hypothesis2\textgreater{}2nd-Best Prediction \textless{}/hypothesis2\textgreater\\ ...\\ \textless{}hypothesisN\textgreater{}Nth-Best Prediction \textless{}/hypothesisN\textgreater\\ =\textgreater{}Correct Transcription: Label 2\\ ...\\ ExampleK:\\ Predicted Transcriptions:\\ \textless{}hypothesis1\textgreater{}1st-Best Prediction \textless{}/hypothesis1\textgreater\\ \textless{}hypothesis3\textgreater{}2nd-Best Prediction \textless{}/hypothesis2\textgreater\\ ...\\ \textless{}hypothesisN\textgreater{}Nth-Best Prediction \textless{}/hypothesisN\textgreater\\ =\textgreater{}Correct Transcription: Label K\\ Query:\\ Predicted Transcriptions:\\ \textless{}hypothesis1\textgreater{}1st-Best Prediction \textless{}/hypothesis1\textgreater\\ \textless{}hypothesis3\textgreater{}2nd-Best Prediction \textless{}/hypothesis2\textgreater\\ ...\\ \textless{}hypothesisN\textgreater{}Nth-Best Prediction \textless{}/hypothesisN\textgreater\\ =\textgreater{}Correct Transcription:\end{tabular}} \\ \hline
\textbf{Assistant:}                 & \multicolumn{2}{l|}{Corrected Transcription for the Query’s Predicted Transcription(s)}                                                                                                                                                                                                                       \\ \hline
\end{tabular}%
}
\label{tab:role_content}
\end{table*}

\subsection{Speech-to-Text}

The first step in our pipeline is the conversion of speech data into text. We apply an advanced ASR model to transcribe the audio data. This step is executed on all parts of the dataset, including training, development, and test sets. The initial transcriptions generated by the ASR model serve as the baseline for further error correction. To improve transcription accuracy, beam search decoding with a beam width parameter of 5 is employed. The baseline WER is computed directly from these initial transcriptions.

\subsection{Knowkedge Base}

To enable the retrieval mechanism, we first create a knowledge base of vector representations for all ASR transcriptions in the training set. Each ASR transcription is processed into a TF-IDF vector. The purpose of using TF-IDF is to emphasize word-level features, which are crucial for addressing issues arising from lexical similarities rather than contextual meaning. This knowledge base serves as the foundation for identifying relevant examples during the retrieval phase.

\subsection{Retrieval}

To enhance the error correction process, we employ a retrieval mechanism based on TF-IDF and cosine similarity. For each ASR-transcript in the dev and test sets, we compute a TF-IDF vector representation and compare it against the TF-IDF vectors of all ASR-transcripts in the knowledge base using the cosine similarity metric. The top 5 most similar samples from the knowledge base are retrieved for each dev and test sample. These retrieved samples are then used as in-context examples for the few-shot prompting process.

\subsection{Generation}

In this section, the retrieved samples are used as few-shot examples to construct a prompt for a generative LLM. Each of the 5 selected samples from the training set appears in the prompt as a pair of ground-truth and ASR transcriptions for each sample in the development and test sets. This enables the generative LLM to leverage information from the training data to perform error correction on the target ASR transcription from the test set. The model generates a corrected version of the target text.

\section{Experiments}
\label{sec:Experiments}

\subsection{Setup}

The proposed system leverages the multilingual Whisper-large-v3 ASR model alongside a RAG approach to significantly enhance ASR accuracy. The architecture comprises two primary components: the Whisper ASR model, which delivers robust multilingual speech-to-text conversion, and the RAG model, which enhances transcriptions through retrieval and generation. The Whisper model ensures broad applicability across diverse languages and accents, making it ideal for preprocessing. The RAG model features a TF-IDF-based retriever and a generator based on GPT-4o. The retriever identifies lexically similar passages from a knowledge base, while the generator combines the ASR output with the retrieved passages to produce corrected transcriptions. \textbf{Table \ref{tab:role_content}} details the procedure in which we combine the retrieved information with ASR transcription(s) for error correction.

For experimental evaluation, the CommonVoice-v19 dataset
\cite{ardila2019common} is used as a multilingual benchmark, known for its diverse accents and demographic representation. The training set of this dataset is preprocessed to create a knowledge base for the retriever, enabling it to identify lexically aligned examples. The validation set is utilized for prompt tuning and evaluating intermediate results. The test set is reserved for the final evaluation, where metrics such as Word Error Rate are computed to measure the system’s overall performance. This setup combines a robust architecture and a comprehensive dataset to validate the system’s effectiveness in real-world scenarios.

\subsection{Results}

This section presents the results of our experiments across three scenarios: 1) Focusing on the impact of language-specific text normalization, 2) The \paperabbr\ scenario utilizing a knowledge base, and 3) Enlarging the knowledge base. These approaches are evaluated for their effectiveness in improving the Word Error Rate in Persian ASR systems.

\subsubsection{Impact of Normalization}

In the first scenario, we examined the effect of language-specific normalization on Persian ASR, where significant improvements in Word Error Rate (WER) were observed. For normalization, we utilized the Hazm library, which addresses common text issues in Persian, including: \begin{itemize} \item Correcting spacing and half-spacing errors, \item Unifying various Unicode representations, and \item Converting English numerals to their Persian equivalents. \end{itemize}

Furthermore, we developed a dictionary consisting of approximately 10,000 entries, capturing frequent typing errors and alternative word representations in Persian. This dictionary effectively mapped these variations to their correct forms, thereby enabling more accurate output analysis. The results, summarized in \textbf{Table \ref{tab:persin_baseline}}, demonstrate that normalization led to a notable reduction in WER, with a decrease of approximately 43\% on the development set and 55\% on the test set.
From now on, all results will be based on using normalization.

\begin{table}[htbp]
\centering
\caption{Effect of Language-Specific Normalization on Word Error Rate (WER) in Persian ASR}
\begin{tabular}{ccc}
\toprule
\textbf{} & \textbf{Dev} & \textbf{Test} \\
\hline
Without Normalization & 45.31 & 86.93 \\
With Normalization    & 25.67 & 39.09 \\
\bottomrule
\end{tabular}
\label{tab:persin_baseline}
\end{table}

\subsubsection{The \paperabbr\   Approach} 
In this scenario, we tested the \paperabbr\ approach using a knowledge base consisting of the entire training data.
The
\paperabbr\
approach improves ASR error correction by retrieving relevant examples from the knowledge base based on lexical similarity, using 1-best or 5-best ASR hypotheses to help identify and correct common errors.
The results for this scenario are shown in \textbf{Table \ref{tab:persin_rag}}.

\begin{table}[htbp]
\centering
\caption{Performance of the \paperabbr\ Approach for ASR Error Correction}
\begin{tabular}{cccc}
\toprule
\textbf{} & \textbf{} & \textbf{Dev} & \textbf{Test} \\
\hline
ASR Baseline & & 25.67 & 39.09 \\
\hline
\multirow{2}{*}{1-best} & Vanilla LLM & 17.84 & 32.04 \\
                        & \paperabbr     & 16.28 & 24.29 \\
\hline
\multirow{2}{*}{5-best} & Vanilla LLM & 15.46 & 26.26 \\
                        & \paperabbr     & \textbf{15.39} & \textbf{21.47} \\
\bottomrule
\end{tabular}
\label{tab:persin_rag}
\end{table}

Despite the limited database, the model showed significant improvements over the baseline ASR system, achieving a 40\% improvement on the development set and a 45\% improvement on the test set. Additionally, the use of 5-best hypotheses enriched the knowledge available to GPT-4o, further enhancing its error correction capabilities. These results underscore the importance of a robust and comprehensive knowledge base to maximize the potential of the retrieval-augmented approach.

\subsubsection{Enlarging the knowledge base} 
In the third scenario, we enlarged the retriever's database by utilizing the full CommonVoice dataset. Specifically, we included all audio files from the "Validated" part of the dataset that are not part of the train, development, or test sets, applying a filter to select files with a downvote of 0 and an upvote of at least 2.
This enlargement involved expanding the knowledge base by a factor of 10. 
The results are shown in \textbf{Table \ref{tab:persin_extended}}.

\begin{table}[htbp]
\centering
\caption{Impact of Enlarging the Knowledge Base on \paperabbr\ Approach}
\begin{tabular}{cccc}
\toprule
\textbf{} & \textbf{} & \textbf{Dev} & \textbf{Test} \\
\hline
\multirow{2}{*}{1-best} & \paperabbr & 16.28 & 24.29 \\
                        & \paperabbr\ - Enlarged Knowledge Base & 8.20 & 6.84 \\
\hline
\multirow{2}{*}{5-best} & \paperabbr & 15.39 & 21.47 \\
                        & \paperabbr\ - Enlarged Knowledge Base & 8.47 & 7.22 \\
\bottomrule
\end{tabular}
\label{tab:persin_extended}
\end{table}

\begin{figure*}[htbp]
	\centerline{\includegraphics[width=0.95\linewidth]{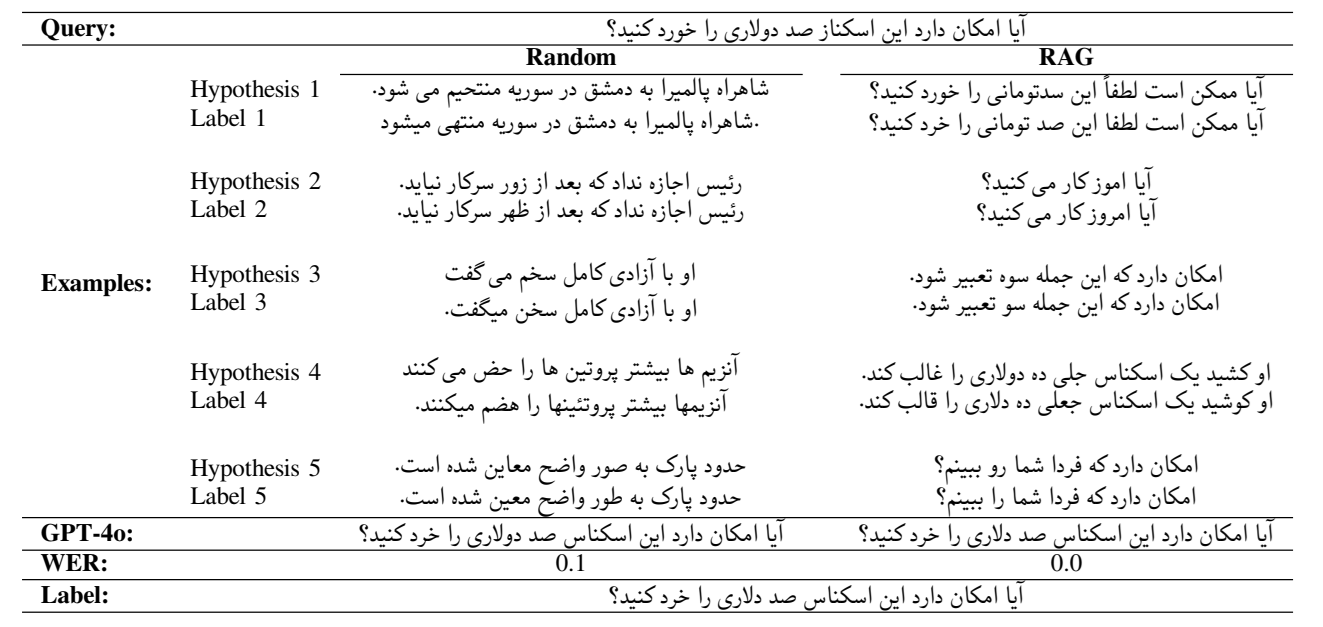}}
	\caption{Showcasing Performance of the Proposed Methodology in 5-Shot 1-Best Scenario.}
	\label{fig2}
\end{figure*}

This enlargement led to significant improvements, achieving a 67\% error reduction compared to the ASR baseline and a 44\% reduction compared to ASR with vanilla GPT on the development set. On the test set, the improvements were even more pronounced, with an \textbf{82\% reduction} compared to the ASR baseline and a \textbf{66\% reduction} compared to ASR with vanilla GPT.

In  \textbf{Figure\ref{fig2}}, we present an example showcasing the effectiveness of RAG compared to the Vanilla LLM (i.e., GPT-4o).

\section{Conclusion}
\label{sec:conclusion}

In this paper, we introduced \paperabbr, a novel approach that integrates RAG with a generative LLM to enhance transcription accuracy, particularly for low-resource languages. We tested our pipeline on Persian, a low-resource language, and demonstrated its effectiveness in reducing ASR errors. The approach utilizes a TF-IDF-based retriever that emphasizes lexical similarity, enabling the system to address ASR errors resulting from phonetic and orthographic similarities. By retrieving lexically similar examples, the system provides relevant context to the generative language model, leading to more accurate corrections.

Additionally, through experimentation, we showed that enlarging the knowledge base improves the retriever's ability to offer relevant examples, which significantly enhances the overall effectiveness of the error correction process. This enlargement plays a crucial role in reducing the WER by providing more accurate corrections.

\paperabbr\ achieves up to an 82\% reduction in WER compared to baseline models, showcasing its potential for low-resource language applications. These results validate the effectiveness of our method, and future work will explore its extension to domain-specific tasks and other linguistic contexts.

\bibliography{ref}

\begin{thebibliography}{10}

\bibitem{errattahi2018automatic}
R.~Errattahi, A.~El~Hannani, and H.~Ouahmane, ``Automatic speech recognition
  errors detection and correction: A review,'' {\em Procedia Computer Science},
  vol.~128, pp.~32--37, 2018.

\bibitem{huang2001spoken}
X.~Huang, A.~Acero, H.-W. Hon, and R.~Reddy, {\em Spoken language processing: A
  guide to theory, algorithm, and system development}.
\newblock Prentice hall PTR, 2001.

\bibitem{baevski2020wav2vec}
A.~Baevski, Y.~Zhou, A.~Mohamed, and M.~Auli, ``wav2vec 2.0: A framework for
  self-supervised learning of speech representations,'' {\em Advances in neural
  information processing systems}, vol.~33, pp.~12449--12460, 2020.

\bibitem{radford2023robust}
A.~Radford, J.~W. Kim, T.~Xu, G.~Brockman, C.~McLeavey, and I.~Sutskever,
  ``Robust speech recognition via large-scale weak supervision,'' in {\em
  International conference on machine learning}, pp.~28492--28518, PMLR, 2023.

\bibitem{toshniwal2018comparison}
S.~Toshniwal, A.~Kannan, C.-C. Chiu, Y.~Wu, T.~N. Sainath, and K.~Livescu, ``A
  comparison of techniques for language model integration in encoder-decoder
  speech recognition,'' in {\em 2018 IEEE spoken language technology workshop
  (SLT)}, pp.~369--375, IEEE, 2018.

\bibitem{shin2019effective}
J.~Shin, Y.~Lee, and K.~Jung, ``Effective sentence scoring method using bert
  for speech recognition,'' in {\em Asian Conference on Machine Learning},
  pp.~1081--1093, PMLR, 2019.

\bibitem{guo2019spelling}
J.~Guo, T.~N. Sainath, and R.~J. Weiss, ``A spelling correction model for
  end-to-end speech recognition,'' in {\em ICASSP 2019-2019 IEEE International
  Conference on Acoustics, Speech and Signal Processing (ICASSP)},
  pp.~5651--5655, IEEE, 2019.

\bibitem{shan2019component}
C.~Shan, C.~Weng, G.~Wang, D.~Su, M.~Luo, D.~Yu, and L.~Xie, ``Component
  fusion: Learning replaceable language model component for end-to-end speech
  recognition system,'' in {\em ICASSP 2019-2019 IEEE International Conference
  on Acoustics, Speech and Signal Processing (ICASSP)}, pp.~5361--5635, IEEE,
  2019.

\bibitem{hrinchuk2020correction}
O.~Hrinchuk, M.~Popova, and B.~Ginsburg, ``Correction of automatic speech
  recognition with transformer sequence-to-sequence model,'' in {\em Icassp
  2020-2020 ieee international conference on acoustics, speech and signal
  processing (icassp)}, pp.~7074--7078, IEEE, 2020.

\bibitem{mcdermott2019density}
E.~McDermott, H.~Sak, and E.~Variani, ``A density ratio approach to language
  model fusion in end-to-end automatic speech recognition,'' in {\em 2019 IEEE
  Automatic Speech Recognition and Understanding Workshop (ASRU)},
  pp.~434--441, IEEE, 2019.

\bibitem{wang2020asr}
H.~Wang, S.~Dong, Y.~Liu, J.~Logan, A.~Agrawal, and Y.~Liu, ``Asr error
  correction with augmented transformer for entity retrieval,'' 2020.

\bibitem{liu2021asr}
X.~Liu, M.~Li, L.~Chen, P.~Wanigasekara, W.~Ruan, H.~Khan, W.~Hamza, and C.~Su,
  ``Asr n-best fusion nets,'' in {\em ICASSP 2021-2021 IEEE International
  Conference on Acoustics, Speech and Signal Processing (ICASSP)},
  pp.~7618--7622, IEEE, 2021.

\bibitem{dutta2022error}
S.~Dutta, S.~Jain, A.~Maheshwari, S.~Pal, G.~Ramakrishnan, and P.~Jyothi,
  ``Error correction in asr using sequence-to-sequence models,'' {\em arXiv
  preprint arXiv:2202.01157}, 2022.

\bibitem{prabhavalkar2023end}
R.~Prabhavalkar, T.~Hori, T.~N. Sainath, R.~Schl{\"u}ter, and S.~Watanabe,
  ``End-to-end speech recognition: A survey,'' {\em IEEE/ACM Transactions on
  Audio, Speech, and Language Processing}, 2023.

\bibitem{ma2023n}
R.~Ma, M.~J. Gales, K.~M. Knill, and M.~Qian, ``N-best t5: Robust asr error
  correction using multiple input hypotheses and constrained decoding space,''
  {\em arXiv preprint arXiv:2303.00456}, 2023.

\bibitem{ma2023can}
R.~Ma, M.~Qian, P.~Manakul, M.~Gales, and K.~Knill, ``Can generative large
  language models perform asr error correction?,'' {\em arXiv preprint
  arXiv:2307.04172}, 2023.

\bibitem{ma2024asr}
R.~Ma, M.~Qian, M.~Gales, and K.~Knill, ``Asr error correction using large
  language models,'' {\em arXiv preprint arXiv:2409.09554}, 2024.

\bibitem{li2024investigating}
S.~Li, C.~Chen, C.~Y. Kwok, C.~Chu, E.~S. Chng, and H.~Kawai, ``Investigating
  asr error correction with large language model and multilingual 1-best
  hypotheses,'' in {\em Proc. Interspeech}, pp.~1315--1319, 2024.

\bibitem{hu2024listen}
Y.~Hu, C.~Chen, C.~Qin, Q.~Zhu, E.~S. Chng, and R.~Li, ``Listen again and
  choose the right answer: A new paradigm for automatic speech recognition with
  large language models,'' {\em arXiv preprint arXiv:2405.10025}, 2024.

\bibitem{li2024rag}
S.~Li, H.~Shang, D.~Wei, J.~Guo, Z.~Li, X.~He, M.~Zhang, and H.~Yang, ``La-rag:
  Enhancing llm-based asr accuracy with retrieval-augmented generation,'' {\em
  arXiv preprint arXiv:2409.08597}, 2024.

\bibitem{pusateri2024retrieval}
E.~Pusateri, A.~Walia, A.~Kashi, B.~Bandyopadhyay, N.~Hyder, S.~Mahinder,
  R.~Anantha, D.~Liu, and S.~Gondala, ``Retrieval augmented correction of named
  entity speech recognition errors,'' {\em arXiv preprint arXiv:2409.06062},
  2024.

\bibitem{ghosh2024failing}
S.~Ghosh, M.~S. Rasooli, M.~Levit, P.~Wang, J.~Xue, D.~Manocha, and J.~Li,
  ``Failing forward: Improving generative error correction for asr with
  synthetic data and retrieval augmentation,'' {\em arXiv preprint
  arXiv:2410.13198}, 2024.

\bibitem{tfidf}
C.~Sammut and G.~I. Webb, eds., {\em TF--IDF}, pp.~986--987.
\newblock Boston, MA: Springer US, 2010.

\bibitem{ardila2019common}
R.~Ardila, M.~Branson, K.~Davis, M.~Henretty, M.~Kohler, J.~Meyer, R.~Morais,
  L.~Saunders, F.~M. Tyers, and G.~Weber, ``Common voice: A
  massively-multilingual speech corpus,'' {\em arXiv preprint
  arXiv:1912.06670}, 2019.

\end{thebibliography}
\bibliographystyle{ieeetr}


\end{document}